# Improved Iterative Techniques to Compensate for Interpolation Distortions

A. ParandehGheibi, M. A. Akhaee, A. Ayremlou, M. A. Rahimian, F. Marvasti[1]

Advanced Communications Research Institute
Sharif University of Technology
Tehran, Iran
[1]marvasti@sharif.edu

*Abstract*—**In this paper a novel hybrid approach for compensating the distortion of any interpolation has been proposed. In this hybrid method, a modular approach was incorporated in an iterative fashion. By using this approach we can get drastic improvement with less computational complexity. The extension of the proposed approach to two dimensions was also studied. Both the simulation results and mathematical analyses confirmed the superiority of the hybrid method. The proposed method was also shown to be robust against additive noise.**

***Index Terms*—iterative approach, interpolation distortion, quadrate latice, modular method**

## 1. Introduction

There are several applications in digital signal processing and communication systems that require the reconstruction of an analog signal from its discrete time samples using D/A converters. Several methods have been introduced in the literature in 1970's and 1980's (for a complete survey of interpolation techniques refer to [1]). Sample-and-Hold (S&H: zero-order-hold) and Linear Interpolation (LI: first-order-hold) were the dominant methods before that time. Today, Spline Polynomial interpolation such as B-splines and Cubic splines are the usual interpolation functions [2-4]. These interpolators create some distortions at the Nyquist rate after low pass filtering, especially when S&H or LI are utilized. The main advantage of these types of interpolators is their simplicity which makes them appropriate for practical use in iterative schemes. There are several methods to compensate for this type of distortion such as inverse $sinc$ filtering, over-sampling, nonlinear adaptive algorithms [5-6], a *modular* method [7], and successive approximation using iterative methods [8-9]. The modular method is compared to the inverse $sinc$ filtering in [7] which shows that by using a few number of modules, the performance of the modular method excels the inverse filtering as far as noise is concerned. Over-sampling is not a practical solution due to its bandwidth requirement. The iterative method [8] outperforms the modular method at the cost of more computations.

We present a hybrid method that combines the benefits of the iterative and the modular methods. The advantages of this hybrid method are fast convergence rate, low complexity and reconstruction delay, and robustness against additive noise. In fact, by using this combined approach, a drastic improvement in signal reconstruction was achieved, with low complexity. We then generalize this hybrid method for 2-D signals, and successfully apply it to the interpolation of actual images. The simulation results confirm the superiority of the proposed hybrid scheme for the interpolation of 2-D signals.

The rest of this paper is organized as follows: Section II briefly describes the background on signal reconstruction techniques including modular and iterative methods as well as the extension of the former to 2-D signals. In section III, we propose a hybrid method by applying the modular method in an iterative framework and prove the convergence of the hybrid method. Noise analysis and sensitivity is also discussed in this section. Simulation results and comparison with other methods are presented in section IV. Finally, section V concludes this paper.

## 2. Preliminaries

### *2.1. Modular Method in 1-D*

In this section we give a brief overview of the modular method [7] that compensates the distortion of common interpolators such as Sample and Hold (S&H) and linear order hold by mixing the sum of cosine waves and then passing them through a lowpass filter.

Let $s(t)$ be an interpolation of samples of $x(t)$. An improved reconstruction of $x(t)$ is given by [7]:

$$\hat{x}(t) = LPF\{\hat{x}_s(t)\}, \qquad (1)$$

where

$$\hat{x}_s(t) = s(t)\left[1 + 2\cos\left(\frac{2\pi t}{T}\right) + \cdots + 2\cos\left(\frac{2N\pi t}{T}\right)\right]. \qquad (2)$$

As $N$ increases, $\hat{x}_s(t)$ converges to its ideal samples of $x(t)$ and thus, $\hat{x}(t)$ converges to $x(t)$.

### *2.2. Modular method in 2-D*

We can extend the 1-D Modular method to 2-D signals. For example for S&H we have:

$$h_{S\&H} = \sum_k \sum_{k'} \delta(t_x - kT)\delta(t_y - kT) \star \left(\prod\left(\frac{t_x}{T}\right)\prod\left(\frac{t_y}{T}\right)\right). \qquad (3)$$

Thus in the 2-D case, the distortion function can be interpreted as

$$sinc(f_xT).sinc(f_yT)\prod_{x,y}\left(\frac{f_{x,y}}{T}\right), \quad (4)$$

in which $\prod_{x,y}\left(\frac{f_{x,y}}{T}\right)$ is a rectangular surface represented as an ideal 2-D LPF. In order to compensate the distortion function, we can add up *sinc* functions in 2-D. Although there are different scenarios depending on the sampling scheme, we just focus on rectangular lattice structure which is common in sampling theory. In this case, as illustrated in Fig. 1, each sample is located on the lattice point. Therefore, the ideal LPF can be obtained by:

$$\sum_{k,k'\in Lattic\ point}\sum sinc(f_xT-k')sinc(f_yT-k)\equiv 1. \quad (5)$$

In the time domain, we have:

$$x_s(t)=s(x,y)\left[1+2\cos\left(\frac{2\pi x}{T_x}\right)+2\cos\left(\frac{2\pi y}{T_y}\right)+4\cos\left(\frac{2\pi x}{T_x}\right)\cos\left(\frac{2\pi y}{T_y}\right)+\cdots\right]. \quad (6)$$

It is clear that as the number of modules increases, the results will converge to its ideal value. We shall see that with a few number of modules we can obtain close approximations.

### 2.3. Iterative Method

The iterative method to compensate for interpolation distortions is given by:

$$x_{k+1}(t)=\lambda Gx(t)+(I-\lambda G)x_k(t), \quad (7)$$

where $\lambda$ is the relaxation parameter that determines the convergence rate and $x_k(t)$ is the $k$-th iteration. In addition, Operator $G=PS$ consists of two operators; $P$ is a band-limiting operator and $S$ is a sampling process, e.g., $S$ can be S&H or LI and $P$ can be a lowpass filter.

## 3. Hybrid Approach

One of the main disadvantages of the traditional iterative method is its low convergence rate, even for the optimum relaxation parameters. Since the modular method outperforms simple low-pass filtering, it can be exploited to improve the convergence rate of the iterative method. In order to combine the modular and iterative methods, we incorporate the modular method in each iteration step as shown in Fig. 2.

We will see in the simulation section that with only one module a phenomenal improvement can be achieved. Below, we shall prove the convergence of the hybrid method for S&H interpolation. The proof for other types of interpolation functions is similar.

*Proof of Convergence for the S&H Interpolation:*

For the $P$ and $S$ operators defined for *S&H*, we can write

$$x_{k+1}(t)=\lambda PSx(t)+x_k(t)-\lambda PSx_k(t), \quad (8)$$

where

$$Sx(t)=\left[1+2\cos\left(\frac{2\pi t}{T}\right)+\cdots\right]\sum_n x(nT)\prod\left(\frac{t-n}{T}-\frac{1}{2}\right), \quad (9)$$

In which $\prod(\frac{t}{T})$ is a rectangular function used for S&H. $x_k(t)$ will converge to $x(t)$ in the limit if we have a contraction, i.e., $\|I-\lambda G\|<1$. This implies [2]

$$\|x_{k+1}-x_k\|<\|x_k-x_{k-1}\|. \quad (10)$$

Substituting (8) in (10), we can get:

$$\|x_k-x_{k-1}-\lambda Gx_k+\lambda Gx_{k-1}\|\leq \|I-\lambda G\|\|x_k-x_{k-1}\|=r\|x_k-x_{k-1}\|, \quad (11)$$

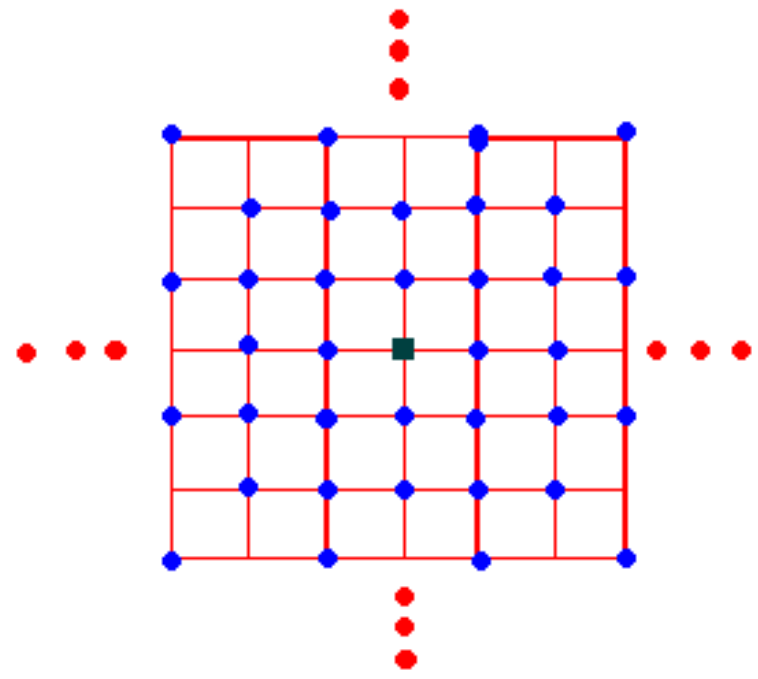

Figure 1. The Quadrature lattice

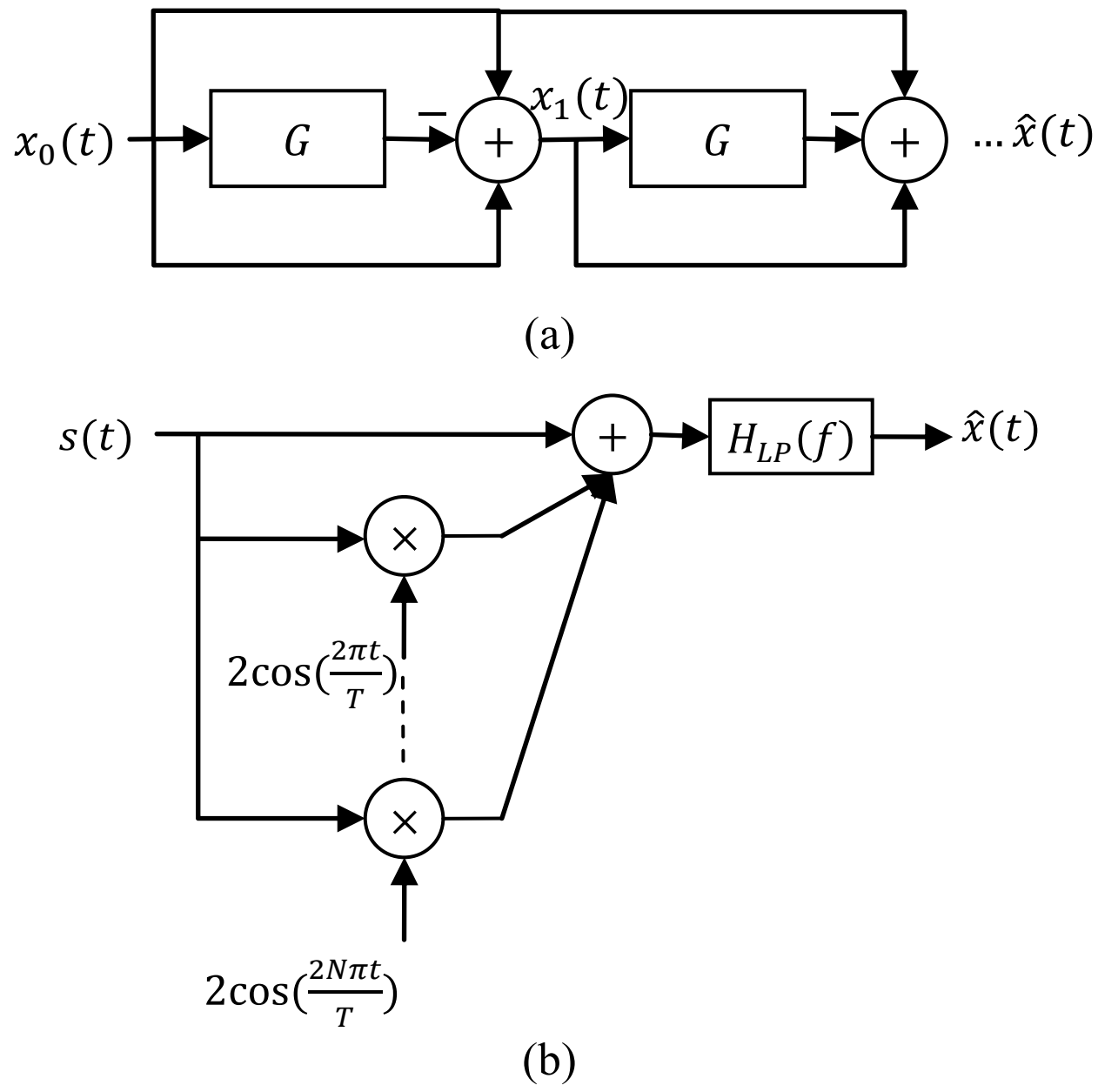

Figure 2. a) The reconstruction block diagram using standard iterative method b) The Modular Method

where $G=PS$ and $0\leq r=\sup\|I-\lambda G\|<1$. Assuming that only one module is applied, the left-hand side of (11) can be rewritten in the frequency domain as follows:

$$\left\|X_k(f)-X_{k-1}(f)-\lambda\prod(fT)([sinc(fT)+sinc(fT-1)+sinc(fT+1)])\times\sum_k\left\{X_k\left(f-\frac{i}{T}\right)-X_{k-1}\left(f-\frac{i}{T}\right)\right\}\right\|, \quad (12)$$

where $\prod(fT)$ is an ideal lowpass filter with the cut-off frequency of $f_c=\frac{1}{2T}$. Assuming that the sampling rate is at the Nyquist rate, (12) becomes

$$\|[X_k(f)-X_{k-1}(f)].\{1-\lambda[sinc(fT)+sinc(fT-1)+sinc(fT+1)]\}\|. \quad (13)$$

Hence,

$$\|X_k(f)-X_{k-1}(f)\|\leq \max|1-\lambda[sinc(fT)+sinc(fT-1)+sinc(fT+1)]|.\|X_k(f)-X_{k-1}(f)\|. \quad (14)$$

To satisfy (14), it is required that

$$0<r=\max|1-\lambda[sinc(fT)+sinc(fT-1)+sinc(fT+1)]|<1. \quad (15)$$

The maximum occurs at $f=\frac{1}{2T}$ and for $\lambda=1$, and we get

$$r=\left|1-sinc\left(\frac{1}{2}\right)-sinc\left(\frac{-1}{2}\right)-sinc\left(\frac{3}{2}\right)\right|=0.06<1. \quad (16)$$

Therefore, the proposed hybrid method converges to the original signal.

From (11), assuming that $G$ is an ideal low pass filter, $r$ can be computed as $r=1-sinc\left(\frac{1}{2}\right)=0.36$. Comparing this value with $r=0.06$ derived for the hybrid method; we expect a drastic convergence rate improvement. For the best convergence rate, the relaxation parameter $\lambda$ should be chosen so that it minimizes $r$, thus at the Nyquist rate, the optimal value for $\lambda$ is given by

$$\lambda_{opt} = \frac{1}{sinc\left(\frac{1}{2}\right) - sinc\left(\frac{-1}{2}\right) - sinc\left(\frac{3}{2}\right)} \cong 0.94. \quad (17)$$

For other types of interpolations, the derivations are similar. For example, for LI we have

$$0 < r = \max|1 - \lambda[sinc^2(fT) + sinc^2(fT - 1) + sinc^2(fT + 1)]| < 1. \quad (18)$$

We get $r = 0.234 < 1$ for $\lambda = 1$, which is less than $r = 0.59$ for the conventional LI interpolation. The optimum value for $\lambda$ is then given by:

$$\lambda_{opt} = \frac{1}{sinc^2\left(\frac{1}{2}\right) + sinc^2\left(-\frac{1}{2}\right) + sinc^2\left(\frac{3}{2}\right)} \cong 1.31. \quad (19)$$

If we mix the signal with more harmonics in each iteration step, as shown in Fig. 2(b), it can be shown that $r$ decreases as we increase the number of modules. In the limit one can write

$$\sum_{k=-\infty}^{\infty} sinc(fT - k) = 1. \quad (20)$$

Hence, $r$ tends to zero for $\lambda = 1$ as the number of harmonics increases, and thus a faster convergence is expected.

### 3.1. *Chebyshev Acceleration of the iterative method*

The iterative and thus the hybrid method can be accelerated by utilizing the two previous iterations based on the Chebyshev acceleration method [11]. Accordingly we have:

$$x_{n>1} = \left(x_1 + x_{n-1} - \frac{2}{A+B} PSx_{n-1} - x_{n-2}\right)\lambda_n + x_{n-2}, \quad (21)$$

where $x_0(t) = \hat{x}(t)$ and $x_1 = \frac{2x_0}{A+B}$. $P$ and $S$ are the operators defined for the iterative method. The constants $A$ and $B$ are frame bounds [11], and should be selected properly for an acceptable performance. There is no unique optimum pair for $A$ and $B$, thus before running the system for the first time they should be selected by experimental methods. The parameter $\lambda_n$ can be calculated as follows:

$$\lambda_n = \left(1 - \frac{\rho^2}{4}\lambda_{n-1}\right)^{-1}, \quad (22)$$

where $\rho$ is defined as $\rho = \frac{B-A}{B+A}$.

The acceleration method improves the iterative method with almost no additional complexity. Notice that the parameter $\lambda$ depends only on the constants $A$ and $B$ and once the $\lambda$ vector is calculated, it is saved in the memory for later implementations.

### 3.2. *Noise Analysis*

Suppose that the proposed hybrid method is used in a noisy environment. For the sake of analysis, white Gaussian noise is added to the original signal before the reconstruction. In this section we will analyze and compare the effects of noise on hybrid and traditional methods. From (8), for the traditional iterative method, we have:

$$x_{k+1}(t) = \lambda Px_s(t) + x_k(t) - \lambda Px_{sk}(t) + (\lambda P)^k n(t), \quad (24)$$

where $n(t)$ is the additive white Gaussian noise to the input, and $x_s(t)$ and $x_{sk}(t)$ are the S&H versions of $x(t)$ and $x_k(t)$, respectively. The necessary constraint on the convergence is the contraction inequality given in (11). Substituting (24) in (10), we obtain

$$\|x_k - x_{k-1} - \lambda G(x_k - x_{k-1}) + ((\lambda P)^k - (\lambda P)^{k-1})n\| \leq r\|x_k - x_{k-1}\|. \quad (25)$$

By invoking the triangle inequality, it is sufficient to have

$$\|x_k - x_{k-1} - \lambda G(x_k - x_{k-1})\| + \|(\lambda P)^k - (\lambda P)^{k-1}\|\|n\| \leq r\|x_k - x_{k-1}\|. \quad (26)$$

If we have a contraction then

$$\|(\lambda P)^k - (\lambda P)^{k-1}\|\|n\| \leq \|\lambda P\|^{k-1}\|n\|. \quad (27)$$

As in the previous section, the following inequality

$$\|X_k - X_{k-1}\|(r - |1 - \lambda sinc(fT)|_{\max}) \geq \|\lambda P\|^{k-1}\|n\| \quad (28)$$

in the frequency domain, should be satisfied for $0 \leq r < 1$. In the worst case we have

$$\|n\| \leq \frac{1}{\lambda^{k-1}}\|X_k - X_{k-1}\|\left(\lambda sinc\left(\frac{1}{2}\right)\right) \cong 0.318\lambda^{2-k}\|X_k - X_{k-1}\|. \quad (29)$$

This implies that as long as the noise standard deviation satisfies (29), the iterations converge. Now, consider the proposed hybrid method. As in (28) we can state that

$$\|X_k - X_{k-1}\|(r - |1 - \lambda[sinc(fT - 1) + sinc(fT) + sinc(fT + 1)]|_{\max}) \geq \lambda^{k-1}\|n\| \quad (30)$$

is a sufficient constraint for the convergence. And for the worst case, we have

$$\|n\| \leq \frac{1}{\lambda^{k-1}}\left(r - \left|1 - \lambda\left[sinc\left(-\frac{1}{2}\right) + sinc\left(\frac{1}{2}\right) + sinc\left(\frac{3}{2}\right)\right]\right|_{\max}\right) \cong 0.531\lambda^{2-k}\|X_k - X_{k-1}\|. \quad (31)$$

Comparing (31) to (30), we conclude that the proposed hybrid method can tolerate more noise power.

### 3.3. *Sampling Rate Analysis*

In the previous sections, the analysis was based on the sampling rate at the Nyquist rate. Suppose the sampling process is $k$ times the Nyquist rate. Invoking the sufficient condition for convergence (11), we have

$$r = \left|1 - sinc\left(\frac{1}{2k}\right) - sinc\left(\frac{1}{2k} - 1\right) - sinc\left(\frac{1}{2k} + 1\right)\right| < 1. \quad (32)$$

For example, for $k = 2$, $r$ is equal to $0.02$. This factor is about 3 times smaller than that of the Nyquist rate; this implies we should expect $10\log(9) = 9.5$ dB improvement in terms of SNR at each iteration step. Similar analysis shows that an equivalent improvement for the LI can be expected. The simulation results presented in section IV confirm the theoretical derivations of this section. Although all the relations were proved for ordinary iterative method, the same relations can be also applied to the Chebyshev accelerated iterative approach [11]. The extension of the proof to the hybrid method is straightforward.

### 3.4. *Modular-Iterative Method in 2-D*

For the proposed method in 2-D, we consider the rectangular lattice sampling process with interpolation function such as S&H and first-order hold. The band limiting process $P$ is an ideal 2-D LPF. All the relations written in the previous section about convergence rate, noise can be restated for the 2-D case. Since the procedure and the mathematical proof closely follow those of the 1-D case, we avoid rewriting it. In the next section, we simulate the performance of the proposed algorithm.

## 4. SIMULATION RESULTS AND DISCUSSION

### 4.1. *Hybrid Method in 1-D*

The performance of different methods discussed so far are evaluated and compared in this section. To have a fair comparison, initial band-limited signals are produced by a Gaussian process with zero mean and an average power of 34 dB. The performance of each method is averaged over 50 signals. The initial signals are FFT lowpass filtered version of pseudo-random signals. During all simulations, we use the same FFT lowpass filter. Parameter $\lambda$ is equal to one and the parameters $A$ and $B$ are set at 1 and 2, respectively. To show the significance of this method, the sampling rate is at the Nyquist rate. The performance criterion for our simulations is the Signal to Noise Ratio (SNR) in dB. To avoid transient errors at the end points, SNR is calculated for interior points and 10% of the end points is ignored. As illustrated in [3], [4] and Fig. 3, the SNR increases monotonically in dB as the number of iterations increases. But it saturates at about 295 dB because of the computer round-off error.

According to the results depicted in Fig. 3, for the classical iterative method, after two iterations, the SNR of about 37 dB is achieved. This means 22dB improvement with respect to simple filtering of sample-and-hold signal. On the other hand, the hybrid method after two iterations reaches 81dB and 99dB for one and two modules, respectively. Hence, the hybrid method improvement is about 66dB for only one harmonic and 84dB for two harmonics, this is quite impressive in real engineering applications.

Fig. 4 shows similar results for the Linear Interpolation (LI). The difference between the hybrid method and the iterative method at the first few iterations is not very significant (about 2-3dB difference). However, as the number of iterations increases, the difference between the two methods becomes apparent. The improvement of the conventional iterative method after eight iterations is about 54dB, while that of the hybrid method is about 89dB. Since the difference in the performance of modular compensators with one and two harmonics is not very significant, only one module has been used in the iteration steps in the following sections.

### 4.1.1. *Effect of the Relaxation Parameter*

For the convergence of the iterative method, $\lambda$ must lie between 0 and 2 [2]. By altering the relaxation parameter, the speed of the convergence rate changes.

Although there are adaptive algorithms to find $\lambda_k$ for the $k$-th iteration step, due to their computational complexity, we would like to find the optimum global $\lambda$ for the best convergence rate; to this end, a quantitative criterion for the rate of convergence is defined. Our criterion is the maximum achievable improvement after 10 iterations divided by the number of iterations (average dB improvement per iteration).

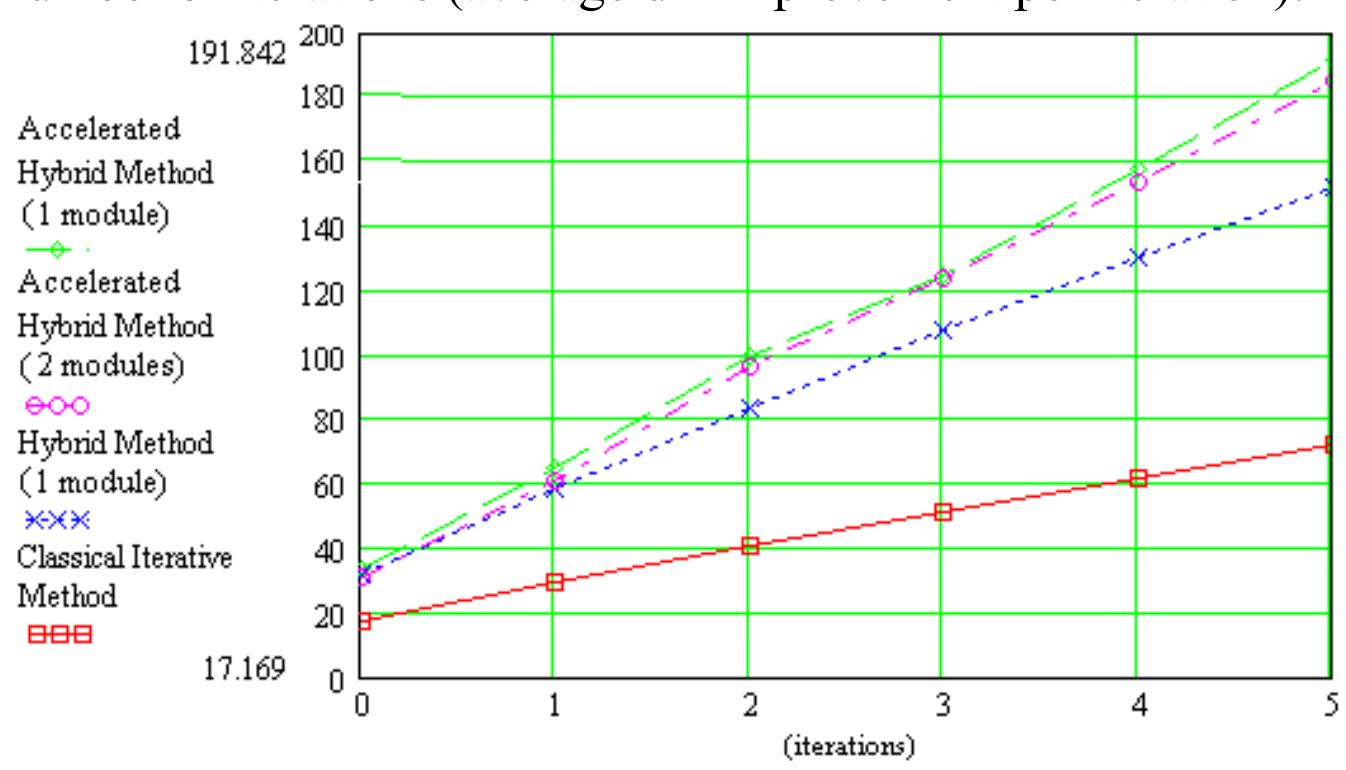

Figure 3. SNR vs. the number of iterations for different methods(Zero-order hold, $\boldsymbol{\lambda} = \mathbf{1}$, 1-D, at the Nyquist rate)

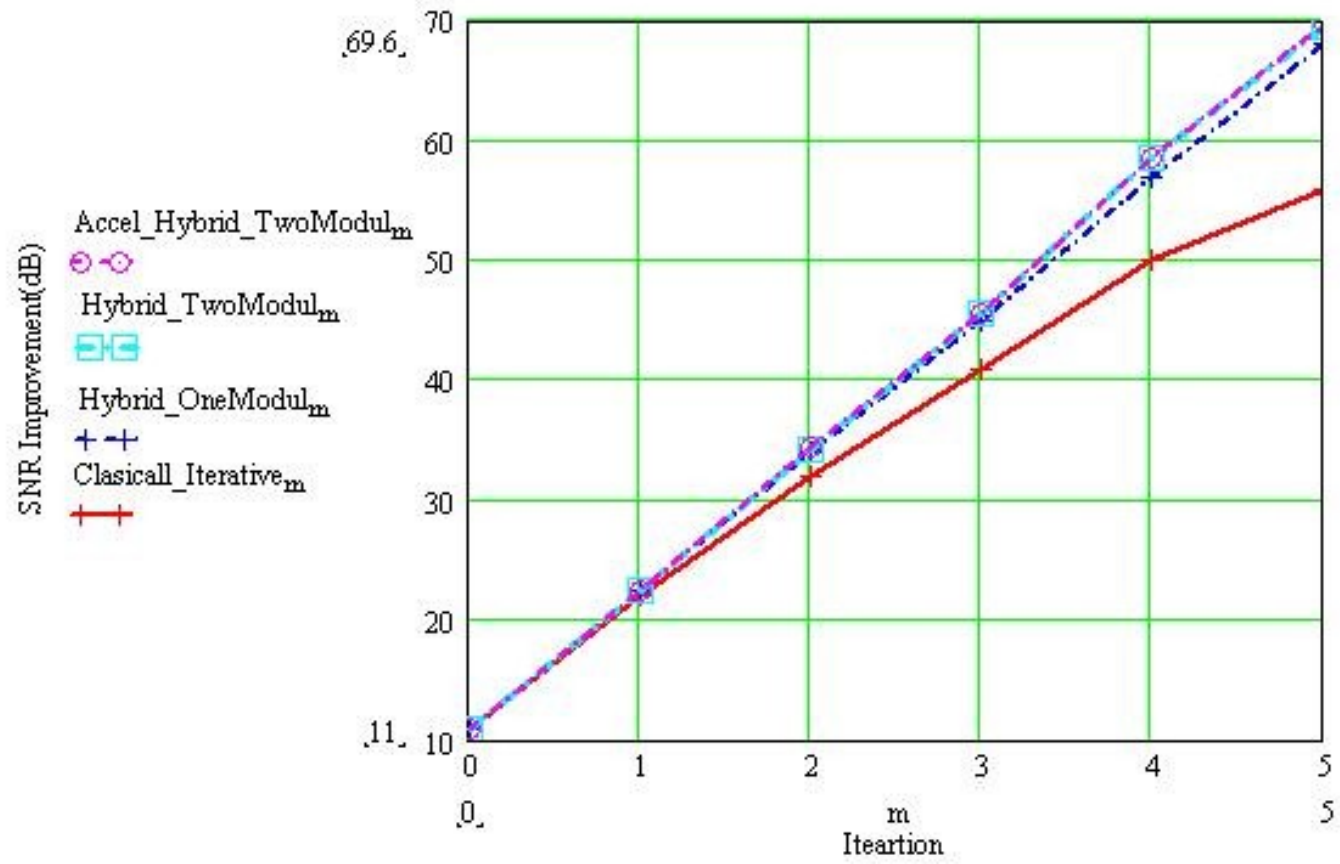

Figure 4. SNR vs. the number of iterations for different methods (First-order hold, $\lambda = 1.3$, 1-D, at the Nyquist Rate)

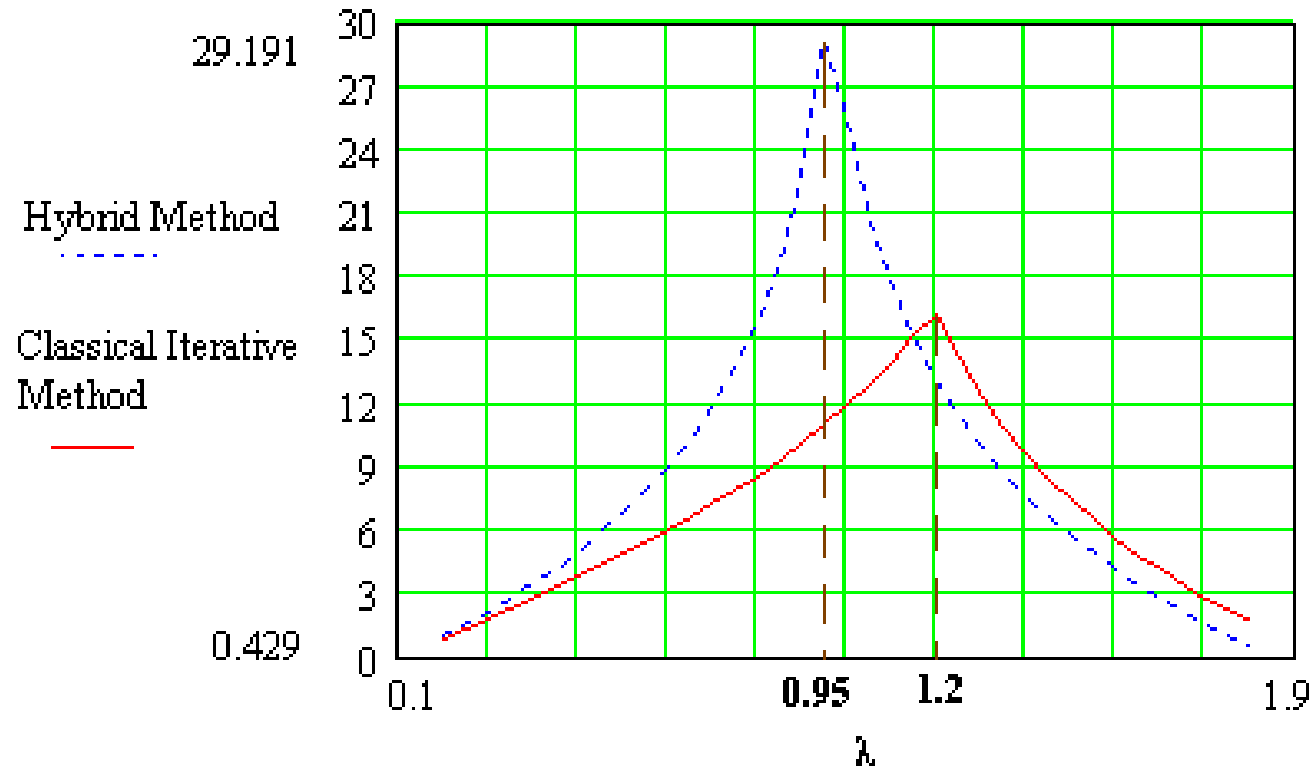

Figure 5. Convergence rate vs. relaxation parameter for different methods for S&H interpolation (1-D, at the Nyquist Rate, with one module).

Fig. 5 shows the convergence rate of the standard iterative method and the proposed hybrid method for different values of the relaxation parameter. We can conclude from this figure that with the best choice of $\lambda$ for each method, our technique outperforms the conventional iterative method by about 13dB per iteration (81%). Also the optimum $\lambda$ is shown to be 0.95 which verifies the theoretical result derived in (17).

The same criterion can be calculated for the LI but the simulation results are omitted for the sake of conciseness; in this case the hybrid method converges 47% faster than the classical iterative method for the optimum relaxation factor. The optimum $\lambda$ is 1.3 which concurs with the theoretical result derived in (19).

### 4.1.2. *Noisy environment*

To study the effect of noise on the convergence rate and maximum achievable SNR, we added a white Gaussian noise, with a power of $-20$ dB, to the signal. This is a model of the channel noise that enters the reconstruction module along with the signal. Fig. 6 shows that after a few iterations, the SNR plot will reach its maximum value. But this value (about 35dB to 45dB) is less than the maximum achievable SNR in the absence of noise (about 300dB). After this climax, the SNR gets worse due to the additive noise and computer round-off errors. Despite the degradations, Fig. 6 shows that the hybrid method for S&H case is still more robust than the conventional method.

### 4.1.3. *The Sampling Rate Effect*

In section III we showed that the SNR and the convergence rate have a direct relationship with the sampling rate. At higher rates the difference between the hybrid and the conventional methods diminishes.

Simulation results show that the SNR, at each iteration step, improves by about 9.5 dB when the sampling rate is doubled. This verifies the relation derived for $r$ in (32), according to which doubling the rate decreases the convergence factor $r$, *by* 3 times and hence $10\log(9) = 9.5\text{dB}$ improvement.

### *4.2. Hybrid Method in 2-D*

Fig. 8 shows the performance of the hybrid approach with different number of modules versus the number of iterations at the Nyquist rate. In fact, zero-order module means a traditional iterative approach. Here, $\lambda$ is set to one, which is the typical choice in most applications.

As it can be observed, the number of modules directly affects the quality of reconstruction. When a small number of modules are used, the iterative method attains a better performance. However, as the number of modules is increased, the role of the iterative method becomes less significant.

Lastly, in order to investigate the performance of the proposed approach in conjunction with the Chebyshev acceleration method, the SNR improvement for the hybrid and accelerated hybrid method is plotted in Fig. 9. For convenience, the proposed method with just zero, and four modules are depicted. As expected, the results show that the accelerated hybrid method outperforms the Chebyshev acceleration iterative approach and the traditional iterative one.

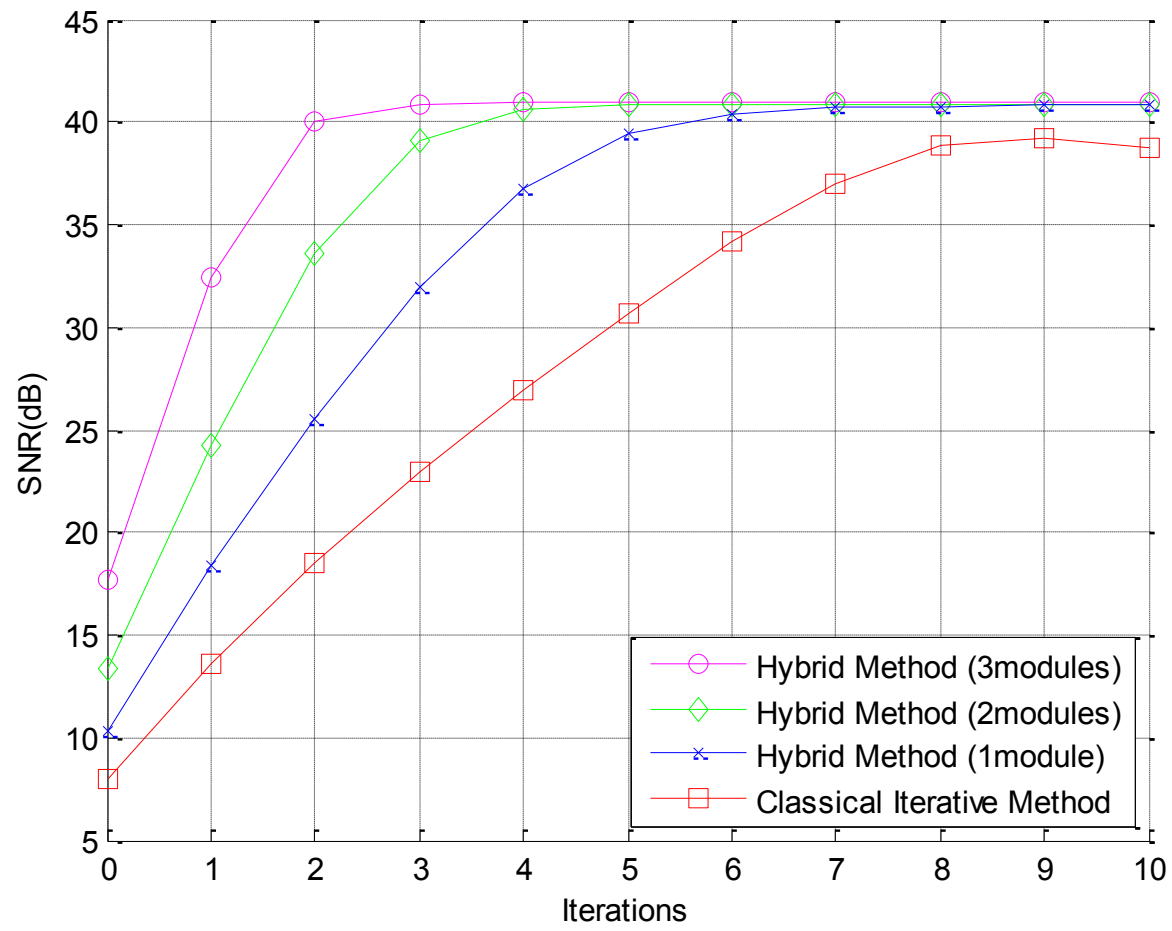

Figure 6. SNR of reconstruction for a random 1-D signal vs. number of iterations for different methods in the presence of noise (S&H, $\lambda = 1$ at the Nyquist rate, the initial S&H/N = 40.38dB).

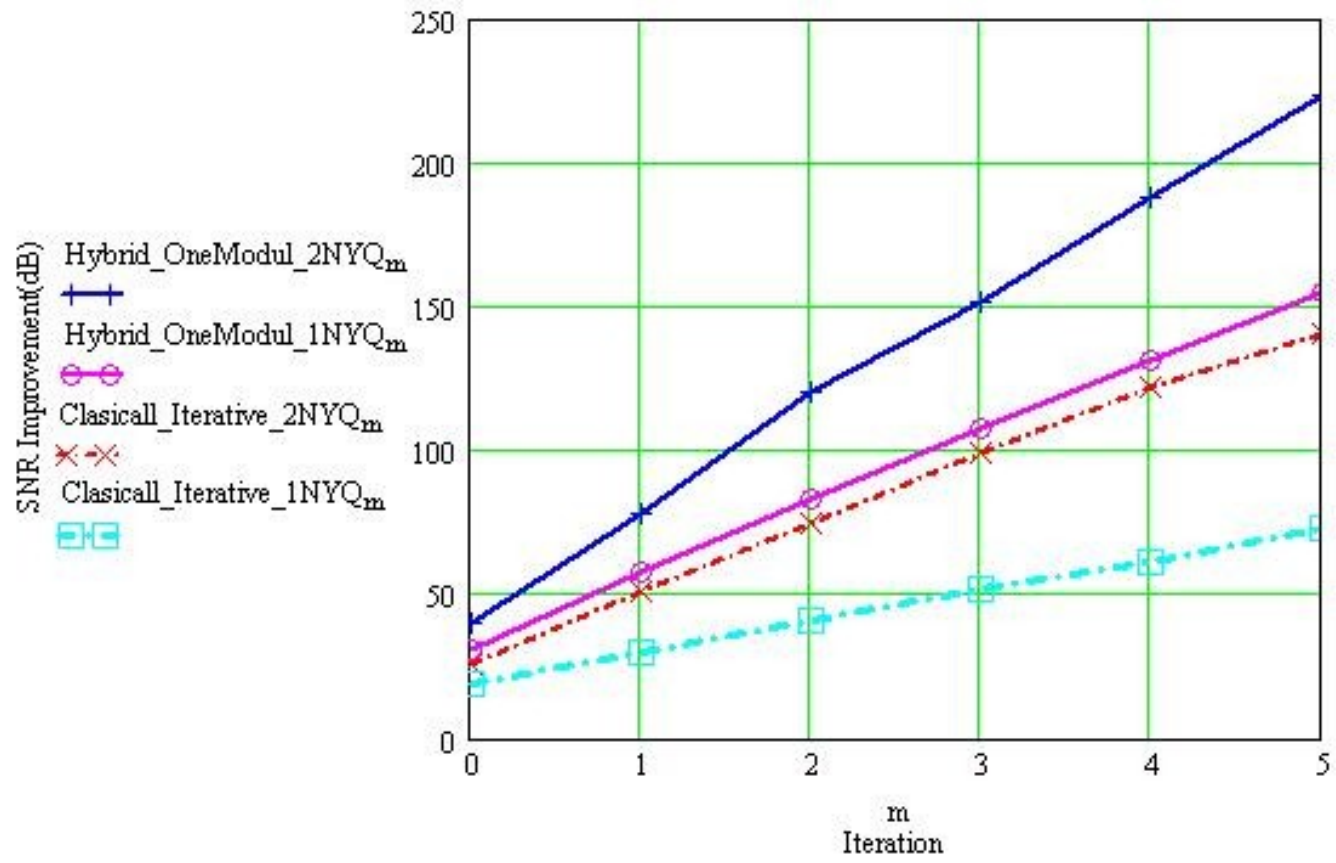

Figure 7. SNR vs. number of iterations. Comparison of the operation of the two methods at twice the Nyquist rate and at the Nyquist rate (S&H, $\lambda = 1$).

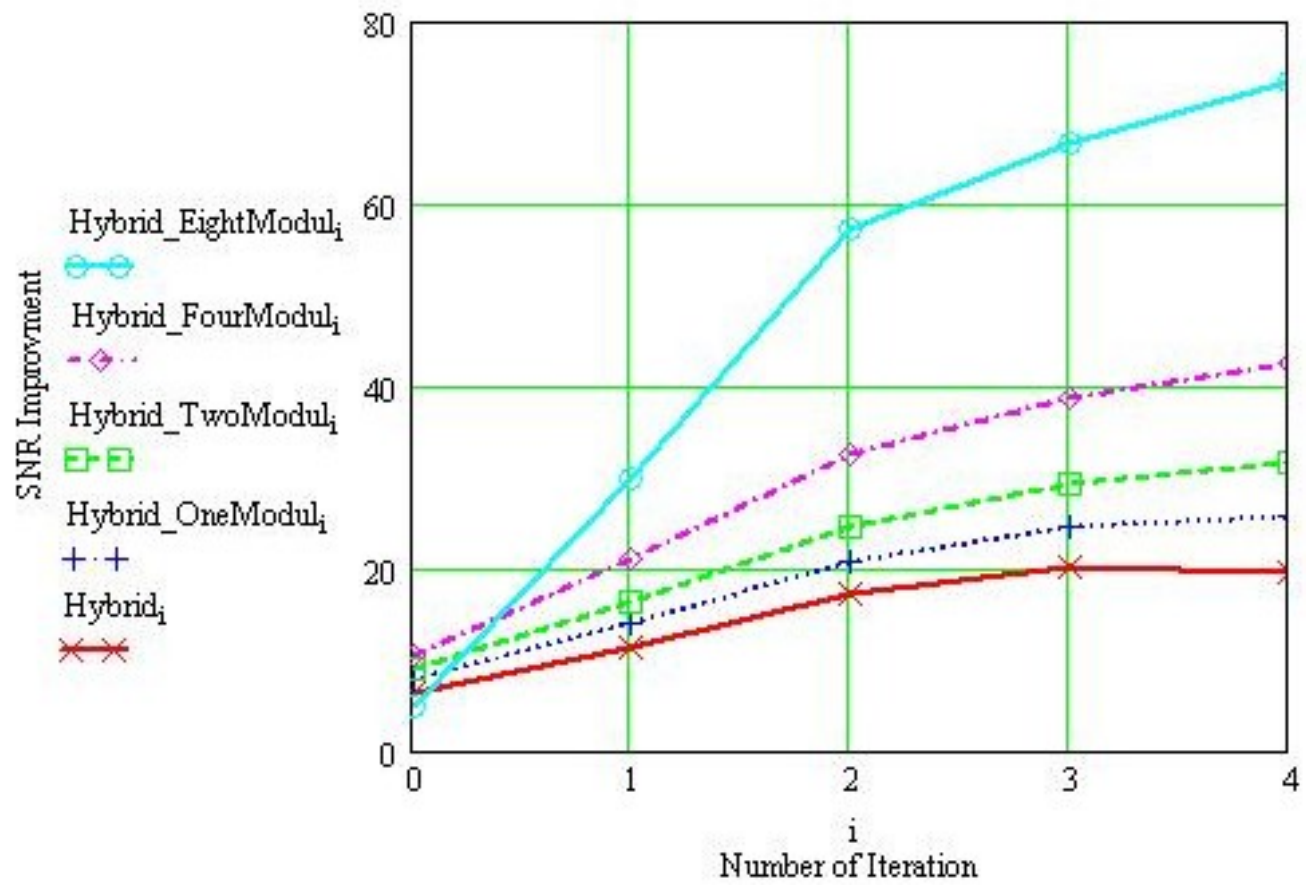

Figure 8. The performance of the hybrid approach with different number of modules versus the number of iterations (2-D, at Nyquist rate).

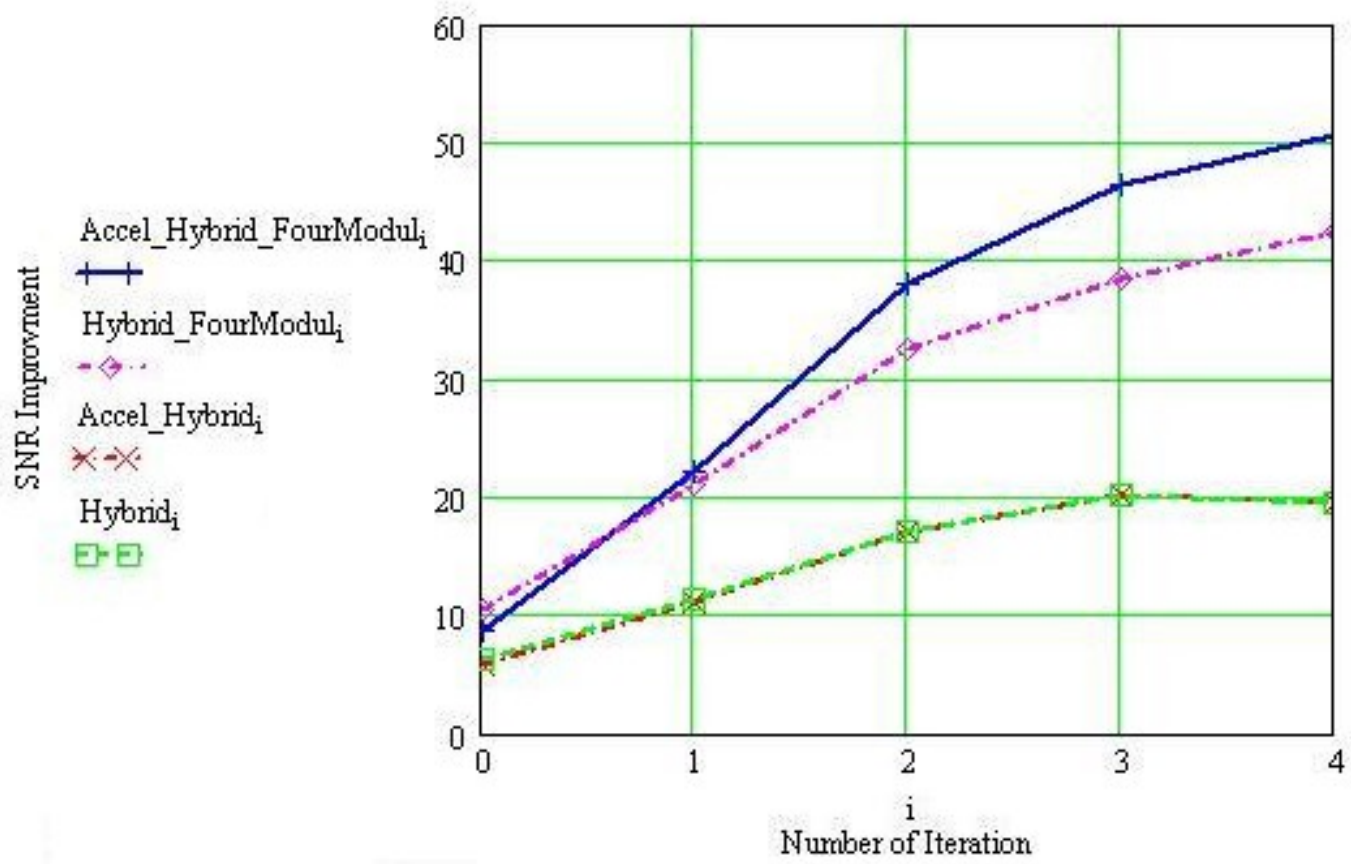

Figure 9. Comparison of the accelerated hybrid approach with the traditional iterative and Chebyshev acceleration iterative methods (2-D, at Nyquist rate).

#### *4.2.1. Effect of the Relaxation Parameter*

In order to evaluate the optimum value of $\lambda$, we calculated the SNR improvement during 5 iterations in the proposed combinational approach with different number of modules. Fig. 10 demonstrates the average SNR improvement versus different values of $\lambda$ for 0, 1, 2 and 3 modules. Based on the Figure, It can be inferred that by increasing the number of modules, the algorithm becomes nearly independent of $\lambda$. Even without any iteration (the starting point), we have a good reconstruction and the iterative algorithm just attains minor increase in SNR. Thus, the role of the iterative algorithm and its relaxation parameter λ decreases with the increasing number of modules.

On the other hand, when the iterative algorithm has a significant role, the value of λ and its effect on the SNR improvement become noticeable. From Fig. 10, for the simple iterative method (without modules), the optimum λ is 1.15, while using modules, this value tends to one.

*4.2.2. Noisy environment*

Similar to 1-D, to simulate the performance of the algorithm in a noisy environment Additive White Gaussian Noise (AWGN), with the same powers as described in section *IV.A.2*, were added to the input signal. Fig. 11 shows the results. As depicted, the curve of SNR improvement in a noisy environment saturates sooner than the noiseless environment. Nevertheless, the proposed method outperforms the traditional iterative method. Here, we have considered the traditional iterative method (no modules) and the hybrid method with 1, 2 and 4 modules. It is evident from this figure that the more the number of modules, the better the reconstruction. Moreover, it can be deduced that, as in the 1-D case, the proposed algorithm enjoys a greater degree of robustness compared to the simple iterative approach.

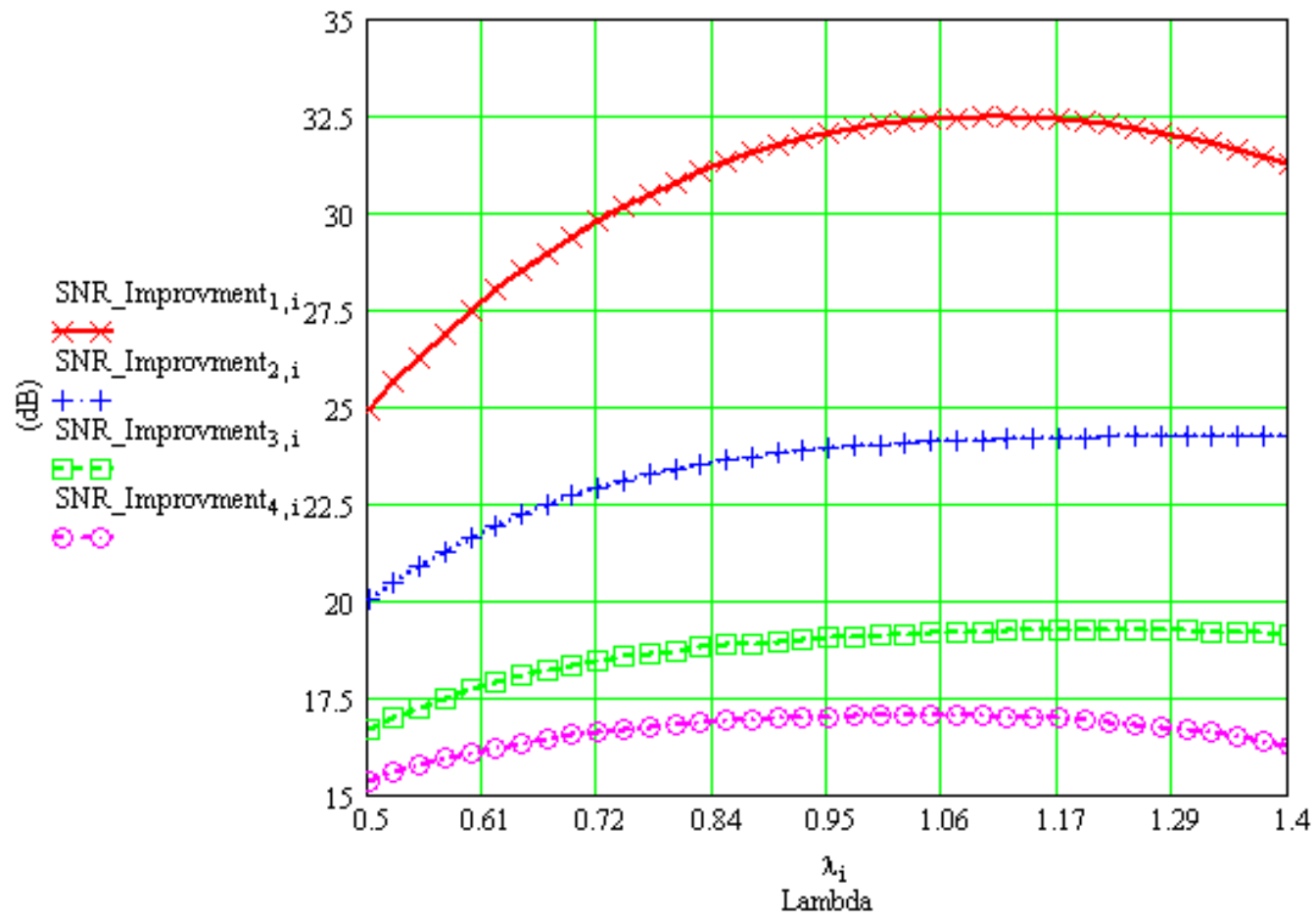

Figure 10. Average SNR improvements for hybrid approach with 0(index $i = 1$), 1($i = 2$), 2($i = 3$) and 3($i = 4$) modules vs. different $\lambda$ (2-D at Nyquist rate).

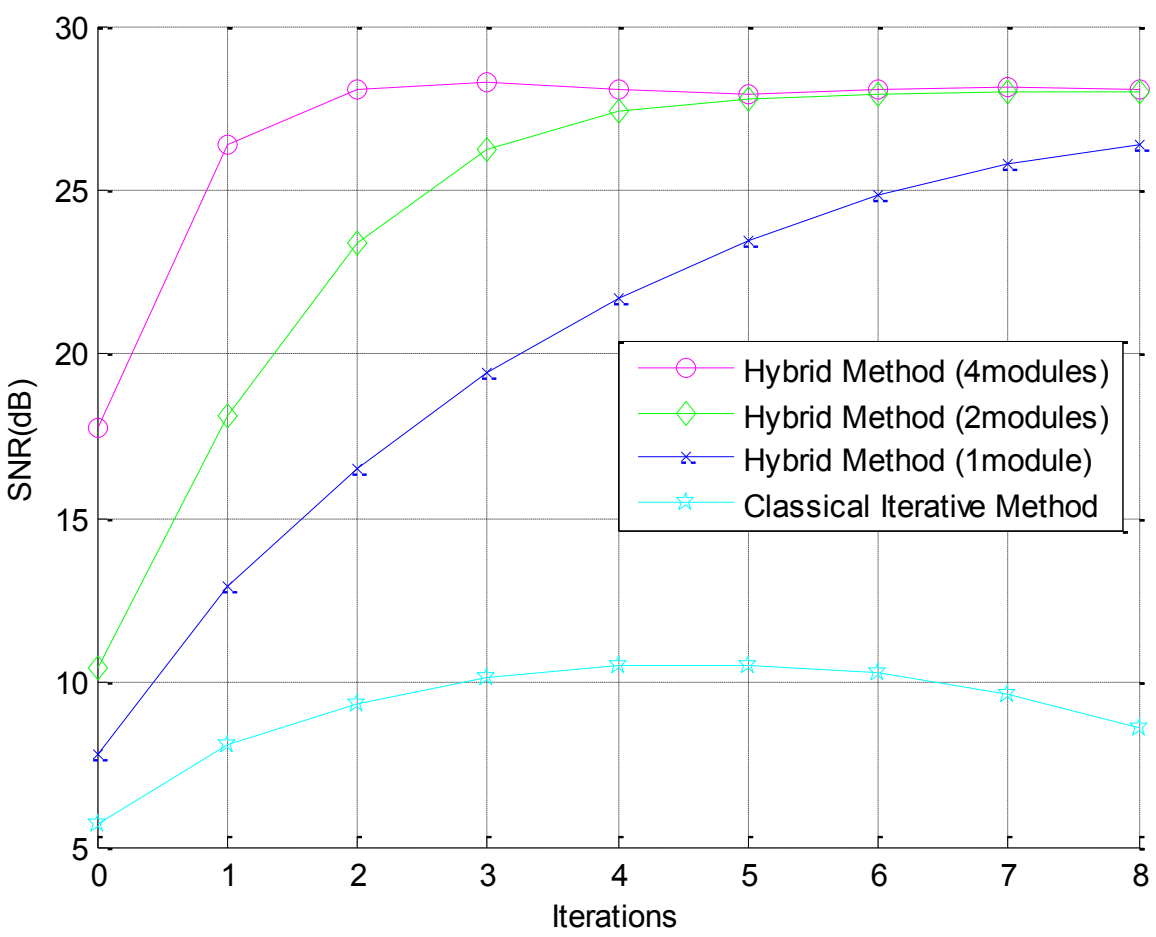

Figure 11. The SNR improvement versus the number of iterationsin the presence of noise, for different methods and with different number of modules, (2-D, the initial S&H/N = 27.69dB).

*4.3. Computational Complexity*

The major advantage of the proposed hybrid method is its higher rate of convergence with less overall computational complexity. The conventional iterative method requires $M(4\log(2N)+2)$ real additions and $M(2\log(2N)+1)$ real multiplications per sample, where $M$ is the number of iterations and $N$ is the FFT block size. But the accelerated hybrid method with one module requires $M(4\log(2N)+4)$ additions and $M(2\log(2N)+3)$ multiplications per sample.

As for the 2-D case, since each of the above computations is performed in one dimension (per each row) and then the same is repeated in the other (per each column), the overall computational complexity is the same as the 1-D case but multiplied by $2K$, where $K$ is the size of the 2-D square matrix. Although the number of computations for the hybrid case in each step of iteration is more, with a fewer number of iterations it achieves the same results and thus its overall computational load is considerably less.

*4.4. Application to Real Images*

We have already shown the computational efficiency of our method to be superior to existing methods of image interpolation. In the end, to evaluate the performance of the proposed method subjectively, we apply our technique to a well-known image (*Lena*). In fact, by this algorithm, we intend to increase the size of the images with acceptable quality.

The errors between the high-resolution originals and reconstructed images are expressed in terms of PSNR (Peak Signal to Noise Ratio) values. Table I shows the errors for 2 × enlargement. Objective comparisons based on PSNR are carried out with conventional bilinear and cubic Spline interpolations (we confirmed the results of [15]) as well as state-of-the art wavelet based methods [12]-[15]. A non-wavelet scheme based on edge-directed interpolation [16]-[17] was also considered to provide a comparison with an established method not operating in the wavelet domain. Our results show that the proposed iterative methods outperform the other methods. Besides, the Hybrid method with only 1 module and 2 iterations exhibits almost the same PSNR performance as the classical iterative method with 10 iterations.

Fig. 12 shows the result of subjective comparison with bicubic Spline interpolation for the image *Lena*. Notice the sharpness of the *Lena* image enlarged with the proposed method in Fig. 12 (mid. Left) compared to the bicubic method (mid. Right), especially around lips, hat, and eyes. Overall, the hybrid method yields images that are sharper than the bilinear or iterative method. Furthermore, as it is depicted in Fig. 12, the enlarged image by the proposed method (bottom Left) has lower errors around edges than other one (bottom Right).

## 5. Conclusion

We proposed a hybrid technique based on the iterative and the modular methods to compensate for the distortion that occurs in the interpolation schemes such as S&H and LI. We theoretically proved that the proposed hybrid method converges much faster than the conventional iterative methods. Simulation results also confirmed the enhanced convergence rate. Furthermore, the superior robustness to noise and lower computational complexity of the Hybrid method were confirmed through both simulations and theoretical analysis. Chebyshev acceleration method was exploited to improve the performance of the Hybrid scheme. Finally, we extended our hybrid technique to 2-D signals, and demonstrated its applicability to real world image interpolations. The aforementioned characteristics make the proposed hybrid approach significant from a practical point of view. In the future we plan to focus on the application of the hybrid method to 2-D signals, where hexagonal sampling has been used.

## ACKNOWLEDGMENT

The authors would like to thank H. Azari for his help with this work.

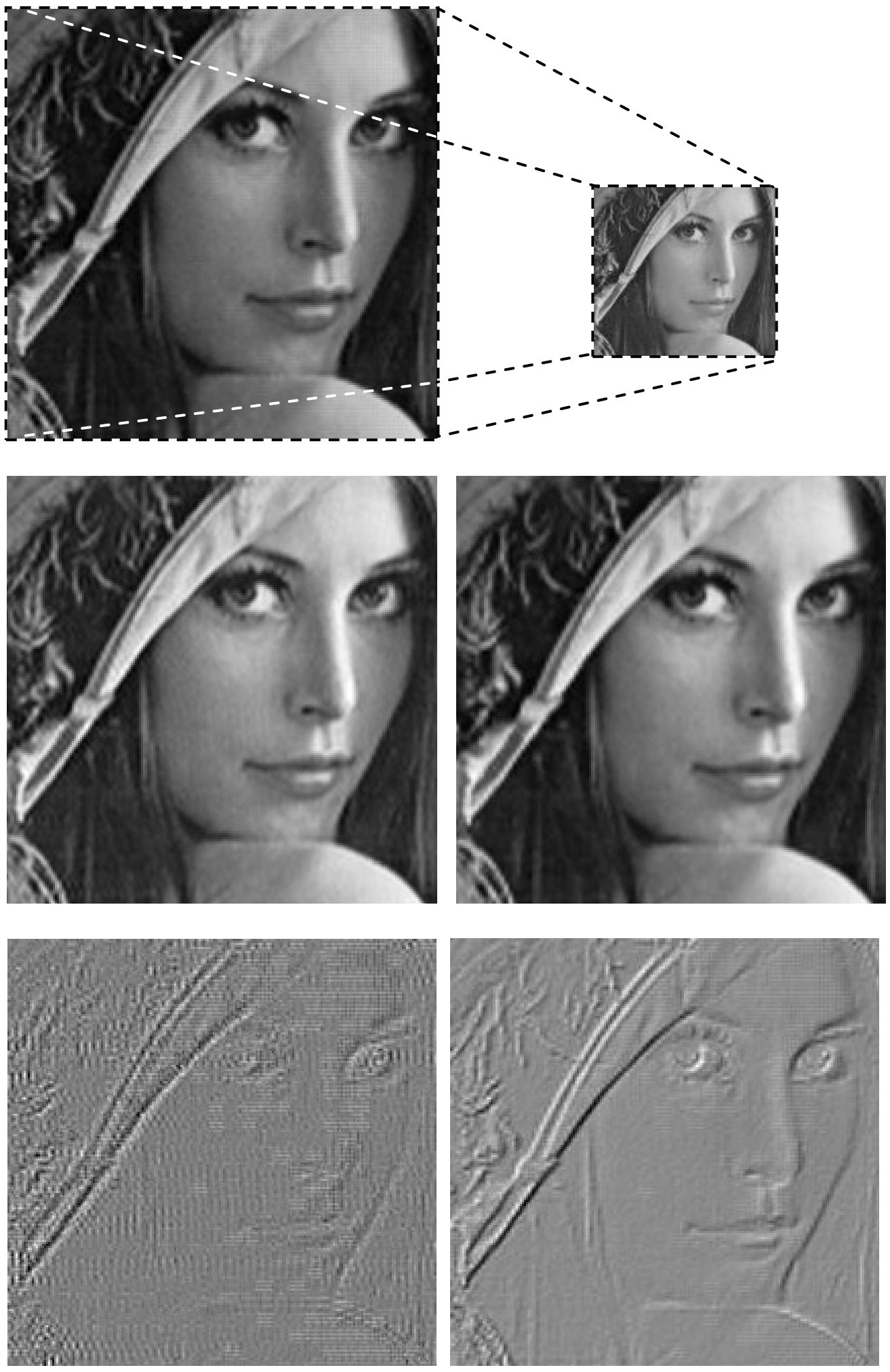

Figure 12. Top Left: Extract from original *Lena* image, Top Right: Original Image reduced by 4, Mid Right: 4× reconstruction using bicubic Spline interpolation, Mid Left: 4× reconstruction using hybrid method with 2 iterations and 1 module, Bottom right: Error corresponding to the bicubic interpolation and Bottom Left: Error corresponding to the hybrid method.

TABLE I
PSNR ($dB$) RESULTS FOR 4× ENLARGEMENT IMAGES (FROM 256×256 TO 512×512)

| Image/Method | *Lena* |
|---|---|
| Bilinear [15] | 30.13 |
| Bicubic [15] | 31.34 |
| NEDI [15-17] | 34.10 |
| WZP –Haar [15] | 31.46 |
| WZP-Db.9/7 [15] | 34.45 |
| Carey et al. [12], [15] | 34.48 |
| HMM [13], [15] | 34.52 |
| HMM-SR [14], [15] | 34.61 |
| WZP – CS [15] | 34.93 |
| SAI [16] | 34.74 |
| **Iterative (2 iter.) [8]** | **35.25** |
| **Iterative (10 iter.) [8]** | **37.39** |
| **Proposed Hybrid (2 iter. And 1 mod.)** | **37.12** |

## REFERENCES

[1] T. H Lehmann, C. Gonner, and K. Spitzer, "Survey: Interpolation Methods in Medical Image Processing", *IEEE Trans. Med. Image*, vol. 18, no. 11, Nov 1999.

[2] H. S. Hou and H. C. Andrews, "Cubic Splines for Image Interpolation and Digital Filtering" , *IEEE Trans. Acoust., Speech, Signal Processing*, vol. ASSP-26, no. 6, pp. 508–517, 1978.

[3] M. Unser, A. Aldroubi, M. Eden, "Fast B-splines Transforms for Continuous Image Representation and Interpolation", *IEEE Trans. Pattern Anal Machine Intell*, vol. 13, no. 3, pp. 277–285, 1991.

[4] P. E. Danielsson and M. Hammerin, "Note: High Accuracy Rotation of Images", *CVGIP: Graph. Models Image Processing*, vol. 54, no. 4, pp.340–344, 1992.

[5] R. R. Schultz and R. L. Stevenson, "A Bayesian Approach to Image Expansion for Improved Definition", *IEEE Trans. Image Processing*, vol.3, no. 3, pp. 233–242, Mar. 1994.

[6] S. Thurnhofer and S. Mitra, "Edge-enhanced Image Zooming" , *Opt. Eng.*, vol. 35, no. 7, pp. 1862–1870, 1996.

[7] F. Marvasti, "A New Method to Compensate for the Sample-and-Hold Distortion", *IEEE Trans. ASSP*, vol ASSP 33, No 3, June 1985.

[8] _____, "An Iterative Method to Compensate for the Interpolation Distortion", *IEEE Trans. ASSP*, vol3, no1, pp 1617-1621, 1989.

[9] _____, "Nonuniform Sampling: Theory and Practice", Kluwer Academic/Plenum Publishers, 2001.

[10] _____, M. Analoui, M. Gamshadzahi, "Recovery of Signals from Nonuniform Samples Using Iterative Methods", *IEEE Trans. ASSP*, vol. 39, Issue 4, pp 872-878, April 1991.

[11] K. Grochenig, "**Acceleration of the Frame Algorithm**", *IEEE Trans. Signal Processing*, vol. 41, pp 3331-3340, Dec 1993.

[12] W.K. Carey, D.B. Chuang and S.S. Hemami, "**Regularity-Preserving Image Interpolation**", *IEEE Trans. Image Processing*, vol. 8, no. 9, pp 1295-1297, Sep. 1999.

[13] K. Kinebuchi, D.D. Muresan and T.W. Parks, "Image Interpolation Using Wavelet-Based Hidden Markov Trees", *Proc. ICASSP '01*, vol. 3, pp. 7-11, May 2001.

[14] S. Zhao, H. Han and S. Peng, "Wavelet Domain HMT-Based Image Superresolution", *IEEE International Conference on Image Proc.*, vol. 2, pp. 933-936, Sep. 2003.

[15] A. Temizel and T.Vlachos, "Wavelet domain image resolution enhancement using cycle-spinning", *IET* **Electronics Letters**, vol. 41, no. 3, pp. 119-121, Feb. 2005.

[16] Xiangjun Zhang and Xiaolin Wu, "Image Interpolation by Adaptive 2-D Autoregressive Modeling and Soft-Decision estimation", IEEE Transactions on Image Processing, Vol. 17, No.6, pp. 887-896, June 2008.

[17] X. Li and M.T. Orchard, "New Edge-directed Interpolation", *IEEE Trans. Image Processing.*, vol. 10, no. 10, pp. 1521-1527, Oct. 2001.